\documentclass[reprint, superscriptaddress, secnumarabic, amssymb, nobibnotes, aps, prl]{revtex4-1}

\usepackage{graphicx}
\usepackage{epstopdf}
\usepackage[T1]{fontenc}
\usepackage[latin9]{inputenc}
\usepackage{amsbsy}
\usepackage{gensymb}
\usepackage[T1]{fontenc}
\usepackage[latin9]{inputenc}
\usepackage{amsmath}
\usepackage{amssymb}
\usepackage{bbm}
\usepackage{physics}
\usepackage{braket}
\usepackage{xcolor}
\allowdisplaybreaks
\usepackage{graphicx}
\usepackage[colorlinks=true]{hyperref}  
\hypersetup{
    bookmarks=true,         
    unicode=false,          
    pdftoolbar=true,        
    pdfmenubar=true,        
    pdffitwindow=false,     
    pdfstartview={FitH},    
    pdftitle={AgSnSe_2},    
    pdfauthor={},     
    pdfsubject={},   
    pdfcreator={},   
    pdfproducer={}, 
    pdfkeywords={} {} {}, 
    pdfnewwindow=true,      
    colorlinks=true,       
    linkcolor=blue, 
    citecolor=blue,        
    filecolor=magenta,      
    urlcolor=blue           
} 
\usepackage[normalem]{ulem}

\begin{document}

\title{\textrm{Superconducting properties of new hexagonal and noncentrosymmetric cubic high entropy alloys}}
\author{K. Motla}
\affiliation{Department of Physics, Indian Institute of Science Education and Research Bhopal, Bhopal, 462066, India}
\author{Arushi}
\affiliation{Department of Physics, Indian Institute of Science Education and Research Bhopal, Bhopal, 462066, India}
\author{S. Jangid}
\affiliation{Department of Physics, Indian Institute of Science Education and Research Bhopal, Bhopal, 462066, India}

\author{P. Meena}
\affiliation{Department of Physics, Indian Institute of Science Education and Research Bhopal, Bhopal, 462066, India}

\author{R. K. Kushwaha}
\affiliation{Department of Physics, Indian Institute of Science Education and Research Bhopal, Bhopal, 462066, India}
\author{R. P. Singh}
\email[]{rpsingh@iiserb.ac.in}
\affiliation{Department of Physics, Indian Institute of Science Education and Research Bhopal, Bhopal, 462066, India}

\begin{abstract} 
Superconducting high-entropy alloys (HEAs) are a newly burgeoning field of unconventional superconductors and raise intriguing questions about the presence of superconductivity in highly disordered systems, which lack regular phonon modes. In our study, we have synthesized and investigated the superconducting characteristics of two new transition elements based HEAs Re$_{0.35} $Os$_{0.35} $Mo$_{0.08} $W$_{0.10} $Zr$_{0.12}$ (ReOMWZ) crystallizing in noncentrosymmetric $\alpha$-Mn structure, and Ru$_{0.35} $Os$_{0.35} $Mo$_{0.10} $W$_{0.10} $Zr$_{0.10}$ (RuOMWZ) crystallizing hexagonal closed-packed structure (hcp). Transition metal-based hexagonal hcp HEA is rare and highly desirable for practical applications due to their high hardness. Bulk magnetization, resistivity, and specific heat measurements confirmed bulk type-II superconductivity in both alloys. Specific heat analysis up to the measured low-temperature range suffices for a BCS explanation. Comparable upper critical fields with the Pauli paramagnetic limit suggest the possibility of unconventional superconductivity in both HEAs.

\end{abstract}

\maketitle

\section{Introduction}
The interest in studying HEAs has grown faster in material science and condensed matter physics over the past few decades \cite{LSun, YF}. HEAs have been adopted as a new and unique concept of metallic alloying in which five or more elements combined between 5 and 35 atomic $\%$ \cite{nano, Yeh, DB, SGuo}. Due to the atomic size difference, these alloys produce more compositional disorder (high mixing entropy) than conventional alloys and are therefore referred to as high entropy alloys (HEA) \cite{corrosionHEA, TNZHT}. Recently, exotic phenomena such as superconductivity and complex magnetism have been exhibited, further enhancing their application spectrum for fundamental and application purposes \cite{prl, fcc, Hexa, Hexagonal, complexmagnetism, boron, NbHea}. However, the superconducting mechanism in these multicomponent alloys is not yet fully understood due to the lack of studies on the two critical ingredients, the electronic band structure and phonon spectra. Recent research also includes the synthesis of new HEA using the combination of different 3d, 4d, and 5d elements, as they provide great tunability for the enhancement of superconducting properties and stabilization in different crystal structures \cite{LSun, prl}. Some HEAs exhibit the unconventional feature of superconductivity, such as the persistence of the zero resistance state on the application of high external pressure, large upper critical field values in HEAs hosted by 4d and 5d transition metals, and the Debye temperature in the elemental range \cite{pressure, Cava, muonHEA}. These extraordinary features increased our curiosity to synthesize new HEAs and explore the exact mechanism inside these disordered materials. Therefore, we have combined 4d and 5d transition elements to synthesize two new Osmium based superconducting HEAs Re$_{0.35}$Os$_{0.35}$Mo$_{0.08}$W$_{0.10}$Zr$_{0.12}$ (ReOMWZ) and Ru$_{0.35}$Os$_{0.35}$Mo$_{0.10}$W$_{0.10}$Zr$_{0.10}$ (RuOMWZ) and investigated their normal and superconducting properties. RuOMWZ crystallized in a hexagonal crystal structure that has been reported for only 1 \% HEA discovered so far, as Miracle $et$ $al.$ Even among hexagonal HEA, those based on transition metals are infrequent yet in high demand for practical applications due to their exceptional hardness compared to body-centered cubic (bcc) and face-centered cubic (fcc) structures \cite {Hexagonal, rare, MF}. By substituting the Ru element with Re, ReOMWZ stabilized in a $\alpha$-Mn non-centrosymmetric (NC) structure which has been reported for some binary Re-based superconductors exhibiting a spontaneous magnetic field on entering the superconducting state viz. time-reversal symmetry breaking; a phenomenon considered as a crucial parameter to exhibit the unconventional superconducting nature in binary NC superconductors \cite{noncentrosymmetric, MoRe, Re6Zr, Re5.5Ta, Re6Hf, Re6Ti}. However, recent findings call into question the role of NC structure \cite{Nb0.5Os0.5, Re, AuBe, NbOs2, TaOs}; therefore, this combined study of centrosymmetric and non-centrosymmetric crystal structure compounds provides a platform to test the hypothesis regarding the role of only the absence of an inversion center and disorder in the origin of unconventional superconductivity in HEAs as well. Moreover, the present study of both compounds will also provide a better understanding of the superconducting mechanism of HEAs based on 4d and 5d transition metals.
\\
In this study, we examined the structure and properties of the superconducting state of both ReOMWZ and RuOMWZ using X-ray diffraction, magnetization, electrical transport, and specific heat measurements, which confirm that ReOMWZ crystallized in the NC $\alpha$ -Mn structure with a superconducting transition temperature T$_{C}$ = 4.90(2) $K$ and RuOMWZ adopted the hexagonal structure with T$_{C}$ = 2.90(2) $K$. The observed upper critical fields exceed the Pauli limiting field, indicating the possibility of unconventionality in both HEAs. Specific heat measurements on both HEAs in the measured temperature range suggest an isotropic gap opening at the superconducting transition. 

\begin{figure} 
\centering
\includegraphics[width=1.0\columnwidth, origin=b]{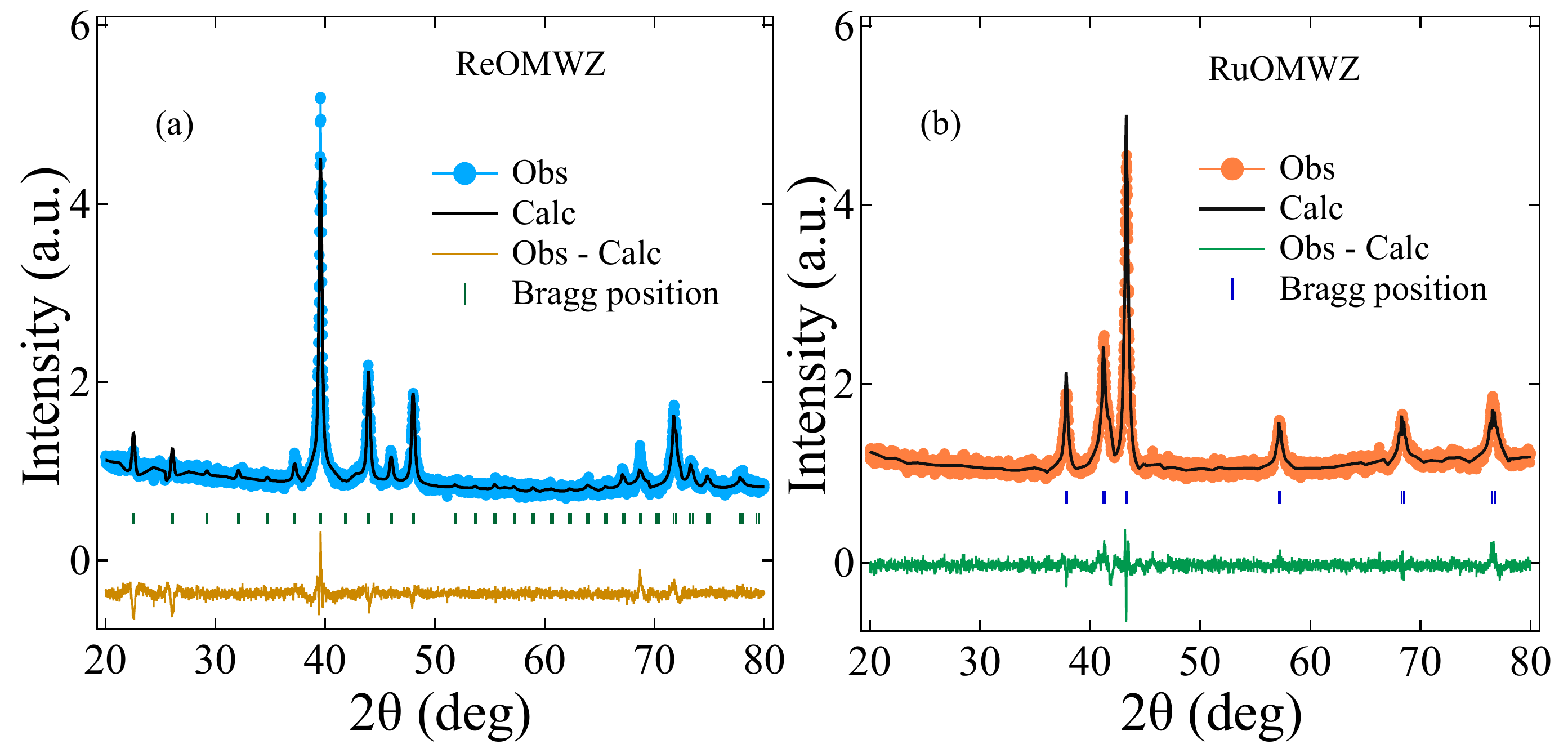}
\caption{The room temperature XRD pattern for (a) ReOMWZ and (b) for RuOMWZ HEAs. The patterns were indexed with $\alpha$-Mn crystal structure for (a) and hexagonal crystal structure for (b).}
\centering
\label{Fig1}
\end{figure}

\section{Experimental Details}
To synthesize polycrystalline samples of ReOMWZ and RuOMWZ, we used arc melting. Stoichiometric ratios of high-purity transition metals (4N) Re, Ru, Os, Mo, W and Zr were arc melted in a chamber filled with argon gas. A Ti getter was melted first to eliminate any remaining oxygen, and the resulting compounds were homogenized by melting the ingot multiple times. Both samples were shiny and had minimal weight loss. We determined the phase purity of the HEAs by analyzing the X-ray diffraction (XRD) pattern of the powdered sample using PANalytical instruments with Cu-K$_{\alpha}$ ($\lambda$ = 1.5405 \text{\AA}) radiation. We analyzed the magnetic properties of both samples using a superconducting quantum interference device ((MPMS 3; Quantum Design, Inc.). The electrical resistivity and specific heat were measured using the Quantum Design physical property measurement system ((PPMS; Quantum Design, Inc.) with standard four-probe geometry and the two-tau time relaxation method.\\

\section{Results and Discussion}
\subsection{Structural Characterization}
The powder XRD patterns of as-cast ReOMWZ and RuOMWZ HEA were collected at room temperature, which is shown in Fig. \ref{Fig1}(a) and (b), respectively. Le-bail analysis of both samples was performed using Fullprof software, which confirms that ReOMWZ and RuOMWZ crystallize in single-phase with $\alpha$-Mn ($I$-43$m$) and hexagonal (P63/$mmc$) crystal structure, respectively which is expected according to Hume Ruthery rule, which provides a relation between valence electron concentration (VEC) and crystal structure \cite{Hume}. VEC for ReOMWZ is 6.81, indicating stability in a body-centered cubic structure, whereas the VEC for RuOMWZ is 7.40, corresponding to groups 7 and 8 of the periodic table and hexagonal structure. The values of the lattice parameters obtained for ReOMWZ and RuOMWZ are listed in Table 1. The broadening of the peaks can be attributed to the disordered nature of HEA samples \cite{prl}. \\

\begin{table}[h!]
\caption{Structure parameters for ReOMWZ and RuOMWZ obtained by Le-bail fitting}
\label{Tab:table1}
\begin{center}
\begingroup
\setlength{\tabcolsep}{3pt}
\begin{tabular}[b]{lcr}\hline
Parameters& ReOMWZ& RuOMWZ\\
\hline
\\[0.5ex]                                  
Structure& $\alpha$-Mn & Hexagonal\\
Space group& $I$-43$m$& P63/${mmc}$  \\
a(\AA)& 9.6605& 2.7449\\
b(\AA)& 9.6605& 2.7449\\
c(\AA)& 9.6605& 4.3793\\
\\[0.5ex]
\hline\hline
\end{tabular}
\par\medskip\footnotesize
\endgroup
\end{center}
\end{table}

\begin{figure}
\centering
\includegraphics[width=1.0\columnwidth]{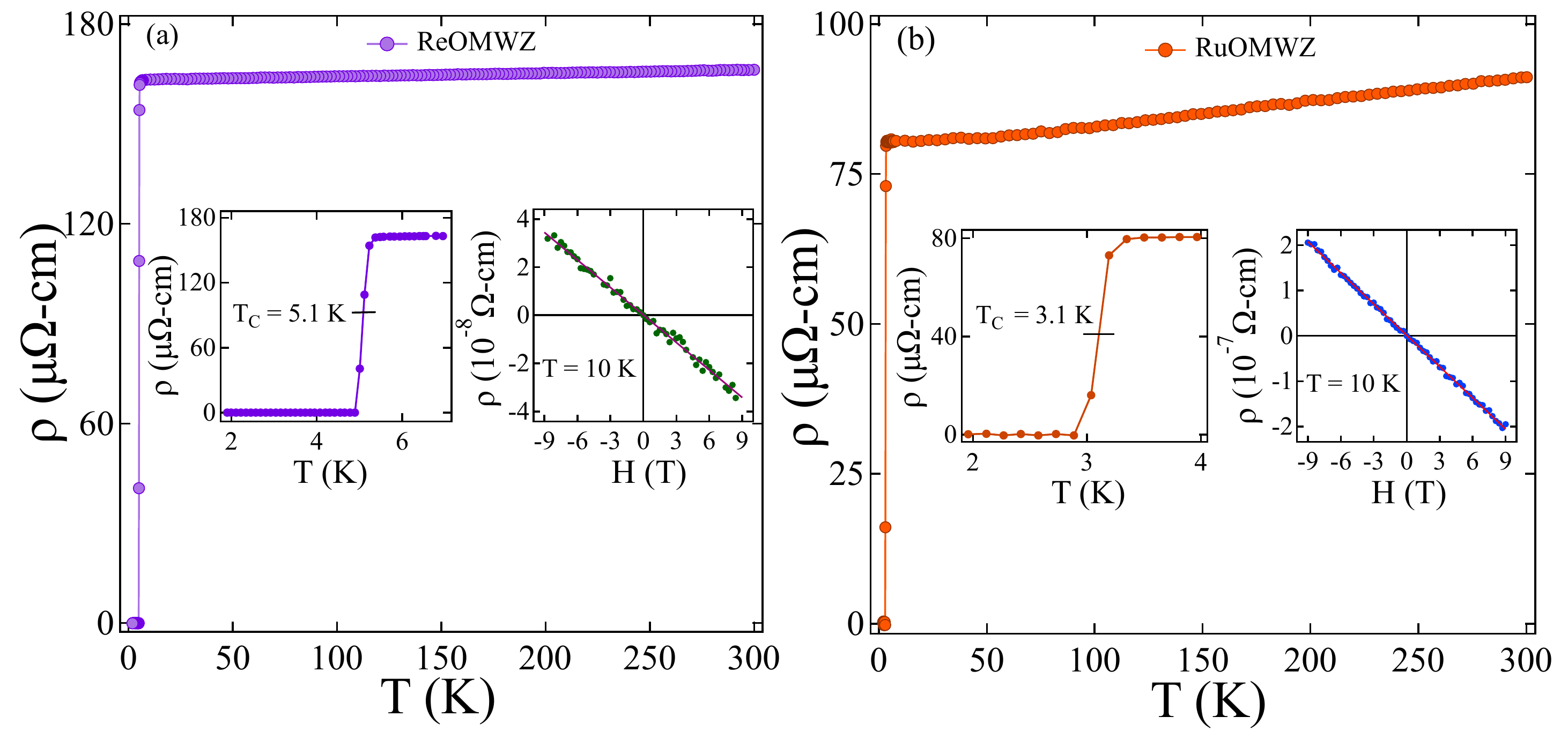}
\caption{ $\rho$(T) measured in the temperature range 1.9 $K$ to 300 $K$ for ReOMWZ and RuOMWZ. Left insets in (a) and (b) shows the sharp T$_{C}$ at 5.1(1) $K$ for ReOMWZ, and 3.1(1) $K$ for RuOMWZ, whereas the right insets represent the Hall resistivity measurements for the respective HEAs.}
\centering
\label{Fig2}
\end{figure}

\subsection {Electrical Resistivity}
We have performed zero-field resistivity measurements in the temperature range of 1.9 $K$ to 300 $K$ for the ReOMWZ and RuOMWZ HEAs, shown in Figs. \ref{Fig2} (a) and (b), respectively. The calculated residual resistivity ratios (RRR) are 1.01 for ReOMWZ and 1.10 for RuOMWZ. The RRRs are very close to 1, which indicates that both HEAs have a poor metallic nature, which is also reported for other HEA superconductors \cite{swavehea, Cava, Complexity, TNZHT}. The expanded view of the zero-field resistivities near the superconducting transition temperature, T$_{C}$ is shown in the inset of Fig. \ref{Fig2} (a) and (b), which exhibit T$_{C}$ at 5.1(1) $K$ and 3.1(1) $K$ for ReOMWZ and RuOMWZ, respectively.
The charge carrier concentration for both samples was calculated using field-dependent ($\pm$9T) resistivity $\rho_{xy}$(H) at 10 $K$ (above T$_{C}$), which is represented in the inset of Fig.\ref{Fig2}(a) and (b). The observed data $\rho_{xy}$ (H) were well fitted using the linear equation and produced slope values of -0.43(1)$\times$ 10$^{-10}$ $\ohm-m T^{-1}$, -2.48(5)$\times$ 10$^{-10}$ $\ohm-m T^{-1}$ for ReOMWZ and RuOMWZ, respectively. The slope values dictate that electrons are the major charge carriers and represent the Hall coefficient, R$_{H}$, which is related to the carrier concentration $n$ using R$_{H}$ = -1/ne, it yields $n$ = (14.5(3), 2.51(5)) $\times$ 10$^{28}$ m$^{-3}$ for ReOMWZ and RuOMWZ, respectively. The range of charge carriers is similar to that reported for other high-entropy alloy superconductors \cite{muonHEA, carrier}.

\begin{figure} 
\centering
\includegraphics[width=1.0\columnwidth, origin=b]{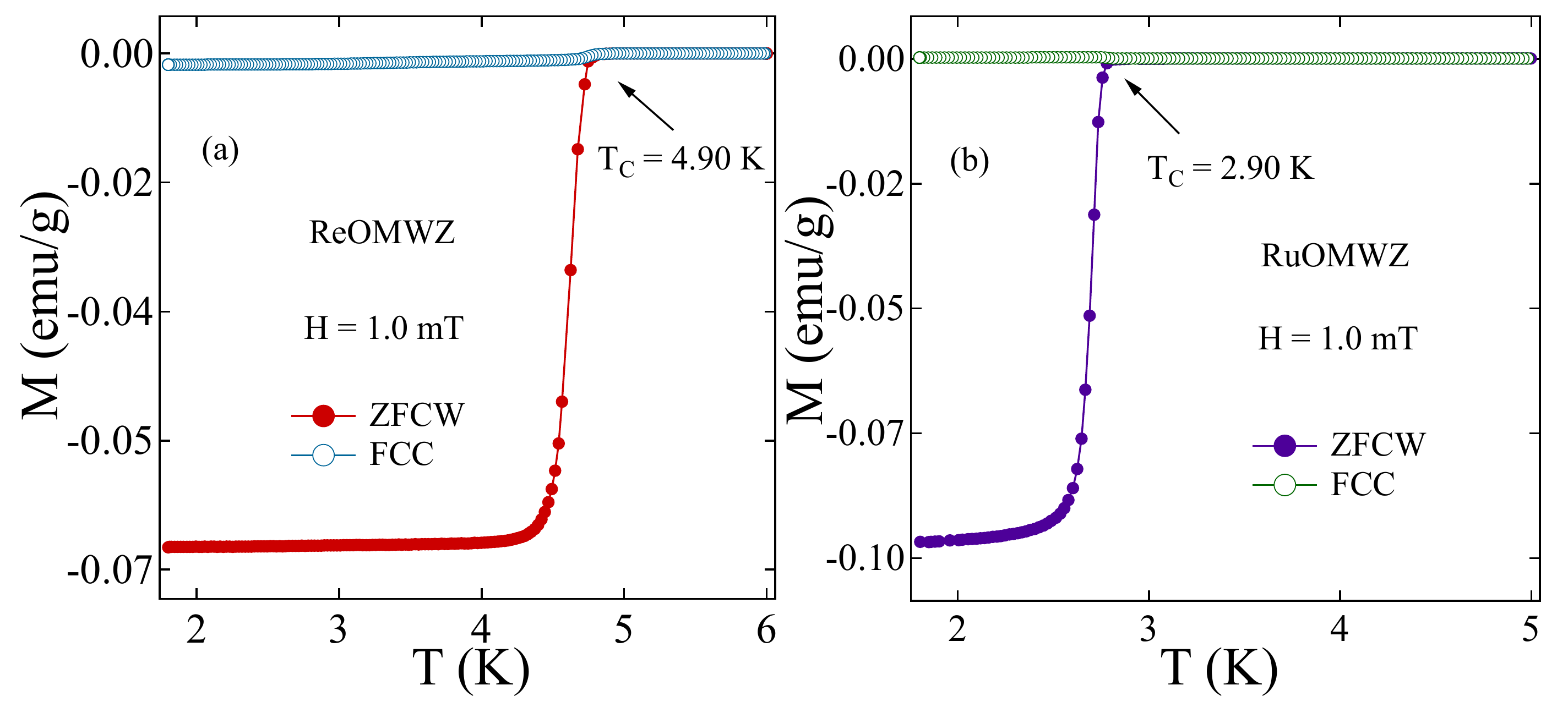}
\caption{DC magnetization data collected in ZFCW and FCC mode under an applied field of 1 $mT$ representing the superconducting transition temperature at 4.90(2) $K$ for ReOMWZ(a) and 2.90(2) $K$ for RuOMWZ(b).}
\label{Fig3}
\end{figure}

\subsection{Magnetization}
\begin{figure} 
\centering
\includegraphics[width=1.0\columnwidth, origin=b]{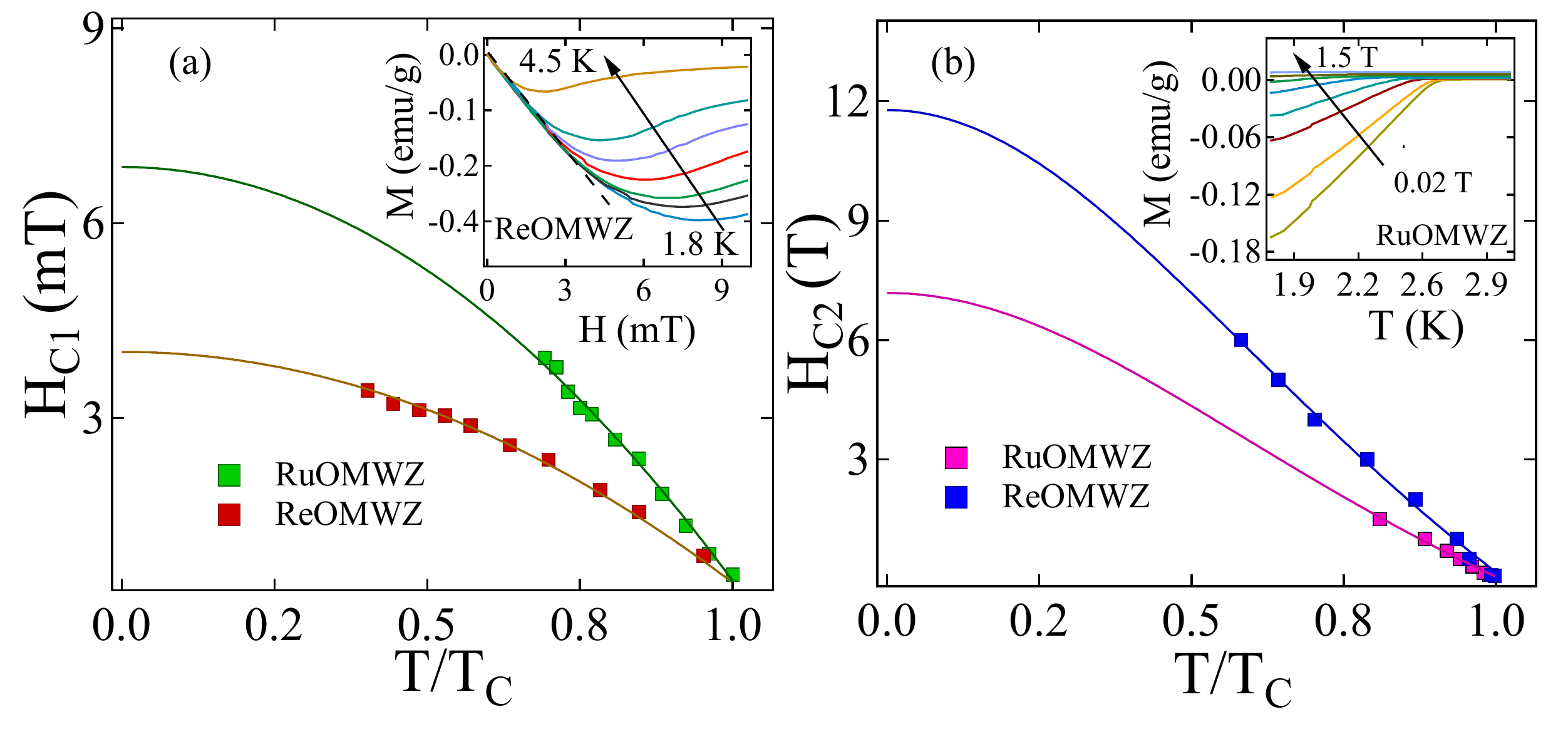}
\caption{(a) Temperature dependence of lower critical field is represented by red and green square symbols for ReOMWZ and RuOMWZ, respectively. The solid lines correspond to the G L Eq. \ref{HC1} fit. Inset in (a) displays the M-H curves for ReOMWZ sample. (b) Variation of upper critical fields w.r.t. temperature for ReOMWZ and RuOMWZ are shown by blue and magenta solid squares, respectively and fitted by G L Eq. \ref{HC2} whereas the inset displays the temperature-dependent magnetization curves under different fields for RuOMWZ.}
\label{Fig4}
\end{figure}

Magnetization measurements confirmed bulk superconductivity in ReOMWZ and RuOMWZ HEA samples in zero field-cooled warming (ZFCW) and field-cooled cooling (FCC) modes under an external magnetic field of 1 $mT$. ZFCW data from both HEA confirm the superconducting nature by exhibiting a strong diamagnetic signal at T$_{C,onset}$ = 4.90(2) $K$ for ReOMWZ and 2.90(2) $K$ for RuOMWZ. The FCC data of both samples have a small diamagnetic signal, which indicates strong pinning.
To evaluate the lower critical field, field-dependent magnetization measurements were performed in the temperature range 1.8 $K$ to T$_{C}$ (ZFCW mode). The magnetization data in a low magnetic field region varies linearly and starts to deviate at a magnetic field value known as the lower critical field, H$_{C1}$. The temperature dependence of H$_{C1}$(T) is evaluated by considering the points deviated from the linear line as shown for ReOMWZ MH data at different T in the inset of Fig. \ref{Fig4}(a), and H$_{C1}$(T) is well explained using the Ginzburg - Landau expression

\begin{equation}
H_{C1}(T)=H_{C1}(0)\left(1-\left(\frac{T}{T_{C}}\right)^{2}\right)
\label{HC1}
\end{equation}
where H$_{C1}$(0) is the lower critical field at 0$K$. The fitting curve estimates H$_{C1}$(0) = 4.04(3) $mT$ for ReOMWZ and 6.86(7) $mT$ for RuOMWZ.
To evaluate the upper critical field value H$_{C2}$(T), the temperature-dependent magnetization M(T) measurements were carried out at different fixed magnetic fields. The M(T) curves shift toward the lower temperatures as the applied field increases and are shown in the inset of Fig. \ref{Fig4}(b).
The onset of diamagnetism was considered a criterion for T$_{C}$. The H$_{C2}$(T) data can be well described with the Ginzburg-Landau model
\begin{equation}
H_{C2}(T) = H_{C2}(0)\left(\frac{1-t^{2}}{1+t^2}\right)
\label{HC2}
\end{equation}\\
where $t$ = T/T$_{C}$ is the reduced temperature and H$_{C2}$(0) is the upper critical field at 0$K$. The extrapolation of theoretical curve up to 0$K$, gives the H$_{C2}$(0) = 11.7(2) $T$ for ReOMWZ and 7.1(2) $T$ for RuOMWZ. 
The value of H$_{C2}$(0) can then be used to estimate the Ginzburg-Landau coherence length $\xi_{GL}$(0) as 
\begin{equation}
H_{C2}(0) = \frac{\Phi_{0}}{2\pi\xi_{GL}^{2}(0)}
\label{xi}
\end{equation}
where $\Phi_0$ is the flux quantum, and substituting the calculated values of H$_{C2}$(0)(from magnetization measurement) yields the coherence length $\xi_{GL}$(0) as 53.0(4) {\AA} and 68.1(9) {\AA} for ReOMWZ and RuOMWZ, respectively.
Another characteristic length parameter, the magnetic penetration depth $\lambda_{GL}$(0) is related to H$_{C1}$(0) via the expression
\begin{equation}
H_{C1}(0) = \frac{\Phi_{0}}{4\pi\lambda_{GL}^2(0)}\left(\mathrm{ln}\frac{\lambda_{GL}(0)}{\xi_{GL}(0)}+0.12\right)   
\label{Lambda}
\end{equation}
where $\Phi_0$ = 2.07$\times$10$^{-15}$T-m$^2$ is the magnetic flux quantum, and using the values of the H$_{C1}$(0) and $\xi_{GL}$(0), we obtained $\lambda_{GL}$ (0) = 4280 {\AA} for ReOMWZ and 3064 {\AA} for RuOMWZ. Using the characteristic length parameters, Ginzburg-Landau categories the superconducting materials as type-I and type-II based on the parameter value, $\kappa_{GL}$ which is expressed as $\kappa_{GL}$=$\frac{\lambda_{GL}(0)}{\xi_{GL}(0)}$. After substituting both values, we calculated $\kappa_{GL}$ = 81 >> ${\frac{1}{\sqrt{2}}}$ for ReOMWZ and 45 >> ${\frac{1}{\sqrt{2}}}$ for RuOMWZ HEA, which indicates strong type II superconductivity in both HEAs. Two mechanisms are responsible for breaking Cooper pairs in type II superconductors; the orbital limiting effect and Pauli paramagnetic limit effect. In the weak limit of BCS superconductivity [see section e], the orbital limiting field H$_{C2}^{orb}$(0) can be estimated via Werthamar-Helfand-Hohenberg (WHH) equation \cite{whh}
\begin{equation}
H_{C2}^{orbital}(0) = - 0.693   T_{C}\left.\frac{dH_{C2}(T)}{dT}\right|_{T=T_{C}}
\label{eqn8:whh}
\end{equation}
using the value of $\frac{dH_{C2}(T)}{dT}$ = - 2.1(1) $T$ for ReOMWZ and -1.60(8) $T$ for  RuOMWZ, we calculated the H$_{C2}^{orb}$(0) as 7.43 $T$ and 3.21 $T$, respectively.
The Pauli-limiting field H$_{P}$ within the BCS limit is defined as 
H$_{c2}^{P}$ = 1.84 $T_{C}$ \cite{pauli}.
By incorporating T$_{C}$ values of both HEAs, H$_{C2}^{P}$(0) was found out to be 9.0 $T$ for ReOMWZ and 5.3 $T$ for the RuOMWZ.
The Maki parameter $\alpha_{M} = \sqrt{2}H_{C2}^{orb}(0)/H_{C2}^{p}(0)$, gives the relative strength of the orbital limiting field to the Pauli paramagnetic limit. The obtained value of $\alpha_{M}$ is 1.15 for ReOMWZ and 0.84 for RuOMWZ. They indicate the influence of the Pauli limiting field over the orbital limiting effect in breaking the Cooper pairs for both HEAs \cite{maki}.

\subsection{Specific Heat}
\begin{figure} 
\centering
\includegraphics[width=1.0\columnwidth, origin=b]{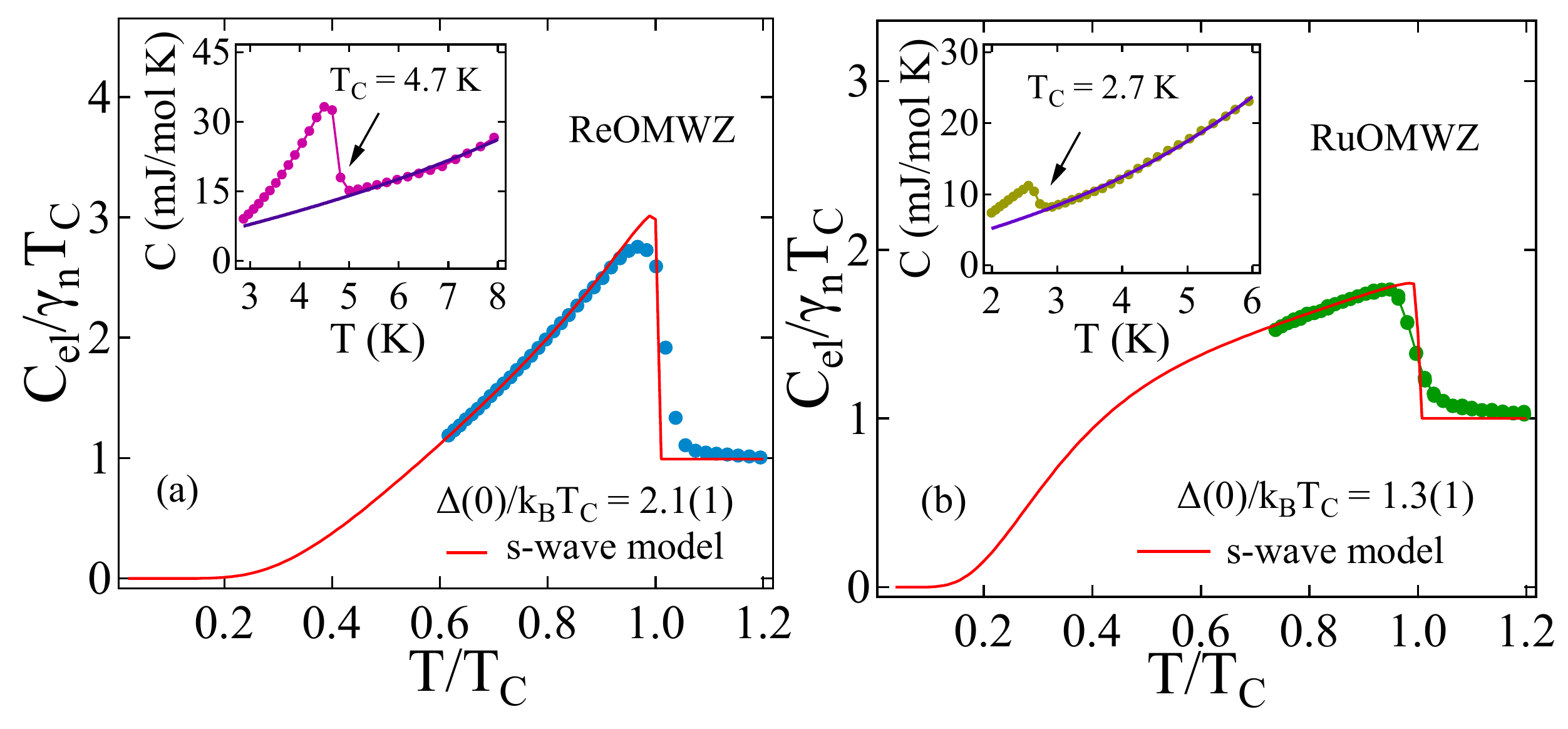}
\caption{Normalized electronic specific heat data at low temperatures for both HEAs were fitted with isotropic BCS s-wave model and shown by the solid lines in (a) and (b). The inset of (a) and (b) shows the fit of the C $vs$ T data using the Eq. \ref{eqn10:C} for ReOMWZ and RuOMWZ, respectively.}
\label{Fig5}
\end{figure}

To characterize the thermal properties of the ReOMWZ and RuOMWZ HEAs, we performed the specific heat measurement in a zero-applied magnetic field, which is shown in the inset of Fig. \ref{Fig5}(a) and (b). A transition is observed from the normal state to the superconducting state showing a jump at 4.70(7) $K$ for ReOMWZ and 2.70(8) $K$ for RuOMWZ, and the observed T$_{C}$ values are in good agreement with the transition from the magnetization and resistivity measurements. Above T$_{C}$, specific heat C $vs$ T data was well explained with the Debye model  
\begin{equation}
C (T) = \gamma_{n} T + \beta_{3}T^{3}
\label{eqn10:C}
\end{equation}
where $\gamma_{n}$ is the electronic contribution or Sommerfeld coefficient and $\beta_{3}$ is the lattice coefficient. Extrapolation of the fit up to 0$K$, estimates $\gamma_{n}$ = (2.52(6), 2.42(1)) $mJ$-$mol^{-1} K^{-2}$ and $\beta_{3}$ = (0.011(1), 0.042(4)) $mJ$-$mol^{-1} K^{-4}$ for ReOMWZ and RuOMWZ, respectively.
The Debye temperature was estimated via the expression
\begin{equation}
\Theta_{D} = \left(\frac{12\pi^{4}RN}{5\beta_3}\right)^{\frac{1}{3}}
\label{eqn11:theta}
\end{equation}
 here R is the universal gas constant (8.31 J mol$^{-1}$ K$^{-1}$), and N = 1 is the number of atom per formula unit. After substituting the parameter values, we get the Debye temperatures as 560(16) $K$ and 357(2) $K$ for ReOMWZ and RuOMWZ, respectively. 
 The Sommerfeld coefficient is directly dependent on the density of state for a non-interactive system as
 \begin{equation}
\gamma_{n} = \left(\frac{\pi^{2}k_{B}^{2}}{3}\right)D_{C}\left(E_{F}\right)
\label{eqn12:gamma}
\end{equation}
where $k_{B}$ is Boltzmann constant (1.38$\times$ $10^{-23} JK^{-1}$), and using the obtained $\gamma_{n}$ values for ReOMWZ and RuOMWZ, the estimated density of state at the Fermi level D$_{C}$(E$_{F}$) are 1.07(2) $\frac{states}{eV f.u}$ for ReOMWZ and 1.03(1) $\frac{states}{eV f.u}$ for RuOMWZ. McMillan proposed a model to estimate the coupling strength between the electron and the phonon $\lambda_{e-ph}$, and the coupling parameter $\lambda_{e-ph}$ is related to $\theta_{D}$, and T$_{C}$ as \cite{McMillan}
\begin{equation}
\lambda_{e-ph} = \frac{1.04+\mu^{*}\mathrm{ln}(\theta_{D}/1.45T_{C})}{(1-0.62\mu^{*})\mathrm{ln}(\theta_{D}/1.45T_{C})-1.04 }
\label{eqn13:ld}
\end{equation}
where $\mu^{*}$ is the screened Coulomb potential, it is taken as 0.13 for the inter-metallic compounds. After substituting the values of $\theta_{D}$ and T$_{C}$, we estimated the electron-phonon coupling parameter $\lambda_{e-ph}$ = 0.53(5) for ReOMWZ and 0.52(3) for RuOMWZ, which suggests that both HEAs are weakly coupled superconductors.
To estimate the superconducting gap values and structure, we evaluate the electronic specific heat at low temperature by subtracting the lattice specific heat from the total specific heat C$_{total}$(T) as 
\begin{equation}
C_{el}(T) = C_{total}(T) - C_{lattice}(T)
\label{Cel}
\end{equation}
Below T$_{C}$, the low temperature electronic specific heat C$_{el}$(T) data of both HEAs was well fitted with isotropic fully gaped BCS model for normalized entropy S, using the expression 
\begin{equation}
\frac{S}{\gamma_{n}T_{C}} = -\frac{6}{\pi^2}\left(\frac{\Delta(0)}{k_{B}T_{C}}\right)\int_{0}^{\infty}[ \textit{f}\ln(f)+(1-f)\ln(1-f)]dy \\
\label{eqn14:s}
\end{equation}
\\
where $\textit{f}$($\xi$) = [exp($\textit{E}$($\xi$)/$k_{B}T$)+1]$^{-1}$ is the Fermi function, $\textit{E}$($\xi$) = $\sqrt{\xi^{2}+\Delta^{2}(t)}$, where $E(\xi $) is the energy of the normal electrons, $\textit{y}$ = $\xi/\Delta(0)$, $\mathit{t = T/T_{C}}$ and $\Delta(t)$ = tanh[1.82(1.018(($\mathit{1/t}$)-1))$^{0.51}$] is the approximated BCS gap value. Below T$_{C}$, the electronic specific heat is related to normalized entropy via the expression

\begin{equation}
\frac{C_{el}}{\gamma_{n}T_{C}} = t\frac{d(S/\gamma_{n}T_{C})}{dt} \\
\label{gap}
\end{equation}
\\
The low-temperature C$_{el}$(T) data was well explained with the isotropic s-wave gap Eq. \ref{gap}, which are shown in Fig. \ref{Fig5}(a) and (b), for ReOMWZ and RuOMWZ, respectively. Due to the base temperature limit (1.8 $K$) of the instrument, C$_{el}$(T) data was fit in the limited temperature range and provided the superconducting gap values $\Delta(0)/k_{B}T_{C}$ = 2.1 (1), 1.3 (1) for ReOMWZ and RuOMWZ, respectively. The superconducting gap value of ReOMWZ indicates strongly coupled superconductivity, while RuOMWZ exhibits weakly coupled BCS superconductivity.

\subsection{Electronic Properties and Uemura Plot}
To evaluate the electronic properties of ReOMWZ and RuOMWZ HEA, we have used some calculated parameters such as the carrier density $n$, the Sommerfeld coefficient $\gamma_{n}$ and the residual resistivity $\rho_{0}$.
The effective mass of quasi-particle m$^{*}$ can be estimated using the expression \cite{Tinkham}
\begin{equation}
\gamma_{n} = \left(\frac{\pi}{3}\right)^{2/3}\frac{k_{B}^{2}m^{*}n^{1/3}}{\hbar^{2}}
\label{eqn16:gf}
\end{equation}
where k$_{B}$ is the Boltzmann constant, and $n$ = (14.5(1), 2.51(1) $\times 10^{28} m^{-3}$) are the carrier density of quasiparticle corresponding to ReOMWZ and RuOMWZ. After substituting the parameter values, we get $m^{*}$ = 3.0(1) m$_{e}$ for ReOMWZ and 4.3(1) m$_{e}$ for RuOMWZ.\\
The Fermi velocity $v_{F}$ is related to m$^{*}$ and $n$ via the expression 
\begin{equation}
n = \frac{1}{3\pi^{2}}\left(\frac{m^{*}v_{F}}{\hbar}\right)^{3}
\label{vF}
\end{equation}
where $\hbar$ is the Planck's constant, using the estimated values of $n$, and m$^{*}$, yields $v_{F}$ = 6.1(3) $\times 10^{5} m/s$ and 2.4(1) $\times 10^{5} m/s$ for ReOMWZ, and RuOMWZ respectively.\\
The mean free path can be estimated using $\rho_{0}$, $m^{*}$ and $v_{F}$, and can be expressed as
\begin{equation}
\textit{l} = \frac{3\pi^{2}{\hbar}^{3}}{e^{2}\rho_{0}m^{*2}v_{\mathrm{F}}^{2}}
\label{l}
\end{equation}
 Substituting the estimated parameter values, we obtained mean free path values for ReOMWZ and RuOMWZ HEAs were determined to be 2.9(3) {\AA} and 18(1) {\AA}, respectively. ReOMWZ value is relatively low and consistent with those observed for other Re-based noncentrosymmetric $\alpha$-Mn structure HEAs \cite{muonHEA}. The low values of the mean free path can be attributed to the high level of disorder caused by the presence of five different elements in the intrinsic disorder of the $\alpha$-Mn structure.
 The coherence length within the BCS model can be expressed in terms of $v_{F}$ as
 \begin{equation}
\xi_{0} = \frac{0.18{\hbar}{v_{F}}}{k_{B}T_{C}}
\label{eqn19:f}
\end{equation}

\begin{figure}
\includegraphics[width=0.95\columnwidth, origin=b]{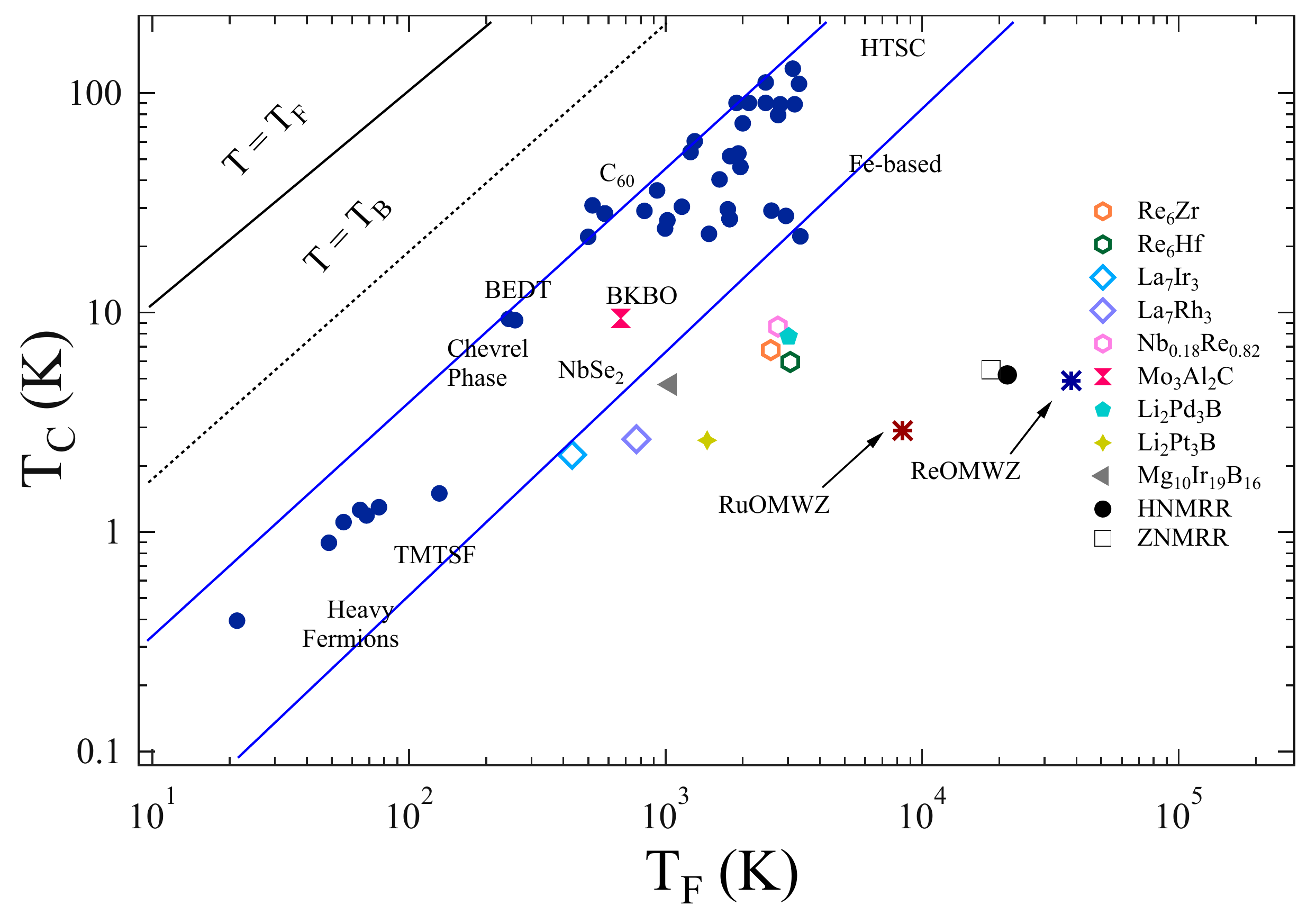}
\caption{The plot between superconducting transition temperature
 $T_C$ versus the Fermi temperature $T_F$ for various superconducting families. The unconventional superconductors are located outside the blue band. ReOMWZ and RuOMWZ, denoted by blue and red stars, respectively, are positioned away from the unconventional band.}
\label{Fig6}
\end{figure}
After using the $v_{F}$ and T$_{C}$ values for ReOMWZ and RuOMWZ, we obtained  $\xi_{0}$ = 1722(92) {\AA} and 1157(56) {\AA}, respectively. The ratio of $\xi_{0}$/$l$ is 580(89) for ReOMWZ and 61(6) for RuOMWZ HEA, which indicates the dirty limit superconductivity in both HEAs and the other estimated parameters of the normal and superconducting state are listed in Table \ref{Tab:table1}.\\
 Uemura $et$ $al$. \cite{U1, U2, U3} provides a classification to differentiate between conventional and unconventional superconductors using the ratio of their superconducting transition temperature T$_{C}$ and the effective Fermi temperature T$_{F}$. If the ratio lies in 0.01 $\leq$ T$_{C}$/T$_{F}$ $\leq$ 0.1, then such materials belong to unconventional superconducting categories in which several strongly correlated exotic superconductors lies, i.e. high T$_{C}$ cuprates, heavy fermions, Chevrel phases, and organic superconductors. However, the compounds whose T$_{C}$/T$_{F}$ $\geq$ 0.1, are considered as the conventional superconductor. The Fermi temperature for the 3D system can be described as follows;
$k_{B}T_{F}$ = ${\frac{{\hbar}^{2}}{2m^{*}} (3{\pi}^{2}n)^{2/3}}$, $n$ is the carrier density which is 14.5(3) $\times 10^{27} m^{-3}$ for ReOMWZ and 2.51(5) $\times 10^{28} m^{-3}$ for RuOMWZ and $m^{*}$ is the effective mass (3.0(1) $m_{e}$for ReOMWZ and 4.3(1) m$_{e}$ for RuOMWZ), which provide T$_{F}$ = 38110(2018) $K$ and 8398(358) $K$ for ReOMWZ and RuOMWZ, respectively. The ratio of T$_{C}$/T$_{F}$ is 0.00012 for ReOMWZ and 0.00034 for RuOMWZ, which places them far away from the unconventional band represented by the solid blue lines in which the unconventional superconductors of different classes reside \cite{JAT, Sundar, Re24Ti5Uemura, Mayoh}.\\    
\begin{table}[h!]
\caption{Superconducting and normal state parameters of ReOMWZ and RuOMWZ HEA}
\begin{center}
\begingroup
\setlength{\tabcolsep}{4pt}
\label{Tab:table1}
\begin{tabular}[b]{lccr}\hline
PARAMETERS& UNITS& ReOMWZ& RuOMWZ\\
\hline
\\[0.5ex]  
VEC& & 6.81& 7.40\\ 
$T_{C}$& $K$& 4.90(2)& 2.90(2)\\
$H_{C1}(0)$& $mT$& 4.04(3)& 6.86(7)\\                       
H$_{C2}^{mag}$(0)& $T$& 11.7(2)& 7.1(2)\\
$H_{C2}^{P}(0)$& $T$& 9.10(3)& 5.39(4)\\
$\xi_{GL}$& {\AA}& 53.0(4)& 68.1(9)\\
$\lambda_{GL}$& {\AA}& 4280(20)& 3064(23)\\
$k_{GL}$& & 81(1)& 45(1)\\
$\Delta(0)/k_{B}T_{C}$& & 2.1(1)& 1.30(1)\\
$m^{*}/m_{e}$& & 3.0(1)& 4.3(1)\\ 
$\xi_{0}/l_{e}$&   & 580(89)& 61(6)\\
$v_{F}$& $10^{5}$ $m s^{-1}$& 6.1(3)& 2.4(1)\\
$n_s$& 10$^{28}$m$^{-3}$& 14.5(3)& 2.51(5)\\
$T_{C}/T_{F}$&   & 0.00012& 0.00034\\
\\[0.5ex]
\hline\hline
\end{tabular}
\par\medskip\footnotesize
\endgroup
\end{center}
\end{table}
 
\section{Conclusion}
In summary, we have synthesized two new Osmium based HEAs Re$_{0.35} $Os$_{0.35} $Mo$_{0.08} $W$_{0.10} $Zr$_{0.12}$ (noncentrosymmetric $\alpha$-Mn) and Ru$_{0.35} $Os$_{0.35} $Mo$_{0.10} $W$_{0.10} $Zr$_{0.10}$ (Hexagonal) which follow the Hume Ruthery rule, relating the valence electron concentration to the crystal structure. Both alloys superconducting and normal state properties were determined using magnetization, resistivity, and specific heat measurements. The bulk superconducting transition temperatures were 4.90(2) $K$ for ReOMWZ and 2.90(2) $K$ for RuOMWZ, with upper critical fields exceeding the Pauli limiting field, suggesting the possibility of unconventionality. Specific heat measurements indicated an isotropic s-wave superconducting gap for both alloys. REOMWZ superconducting parameters are similar to other reported Re-based noncentrosymmetric $\alpha$-Mn HEAs. Further microscopic studies, such as muon spin rotation/relaxation measurements, will help understand crystal structure and disorder role on the superconducting pairing mechanism. Moreover, RuOMWZ HEA shows an exceptional combination of the hcp structure and superconductivity, opening new avenues for investigating cutting-edge superconducting materials. To facilitate its practical applications, a thorough examination of the mechanical properties of the hexagonal RuOMWZ HEA is necessary.

\section{Acknowledgments}
R.~P.~S.\ acknowledges the Science and Engineering Research Board, Government of India, for the CRG/2019/001028 Core Research Grant. Kapil Motla acknowledges the Indian Council of Scientific and Industrial Research (CSIR) Government of India for providing SRF fellowship [Award no; 09/1020(0123)/2017-EMR-I)].


\begin{thebibliography}{References}
\bibitem{LSun} L. Sun, R. J. Cava, Phys. Rev. Materials 3, 090301 (2019).
\bibitem{YF} Y. F. Ye, Q. Wang, J. Lu, C. T. Liu, and Y. Yang, Mat. Today 19, 349 (2016).
\bibitem{nano} J. W. Yeh, S. K. Chen, S. J. Lin, J. Y. Gan, T. S. Chin, T. T. Shun, C. H. Tsau and S. Y. Chang, Adv. Eng. Mater. 6, 299 (2004).
\bibitem{Yeh} J. W. Yeh, S. Y. Chang, Y. D. Hong, S. K. Chen, S. J. Lin, Mater. Chem. Phys. 103, 41 (2007).
\bibitem{DB} D. B. Miracle, and O. N. Senkov, Acta Materialia 122, 448 (2017).
\bibitem{SGuo} S. Guo, and C. T. Liu, Prog. Nat.:Mater. Int. 21, 433 (2011).
\bibitem{corrosionHEA} Y, Qui, S. Thomas, M. A. Gibson, H. L. Fraser, and N. Birbilis, npj. Mater Degrad 1,15 (2017).
\bibitem{TNZHT} S. Vrtnik, P. Ko\v{z}elj, A. Meden, S. Maiti, W. Steurer, M. Feuerbacher, J. Dolin\v{s}ek, Journal of Alloys and Compounds 695, 3530 (2017).
\bibitem{prl} P. Ko\v{z}elj, S. Vrtnik, A. Jelen, S. Jazbec, Z. Jagli\v{c}i\'{c}, S. Maiti, M. Feuerbacher, W. Steurer and J. Dolin\v{s}ek, Phys. Rev. Lett. 113, 107001 (2014).
\bibitem{fcc} N. Ishizu, J. Kitagawa, arXiv 2020, arXiv:2007.00788.
\bibitem{Hexa} B. Liu, J. Wu, Y. Cui, Q. Zhu, G. Xiao, S. Wu, G. Cao and Z. Ren, Scripta Materialia 182 (2020).
\bibitem{Hexagonal}  S. Marik, K. Motla, M. Varghese, K. P. Sajilesh, D. Singh, Y. Breard, P. Boullay, and R. P. Singh, Phys. Rev. Materials 3, 060602 (2019).
\bibitem{complexmagnetism} J. Lu\v{z}nik,P. Ko\v{z}elj, S. Vrtnik, A. Jelen, Z. Jagli\v{c}i\'{c}, A. Meden, M. Feuerbacher, J. Dolin\v{s}ek, Phys. Rev. B 92, 224201 (2015).
\bibitem{boron} K. Motla, Arushi, V. Soni, P. K. Meena, and R.P. Singh, Supercond. Sci. Technol. 35, 074002 (2022).
\bibitem{NbHea} K. Motla, P. K. Meena, Arushi, D. Singh, P. K. Biswas, A. D. Hillier, and R. P. Singh, Phys. Rev. B 105, 144501 (2022).
\bibitem{pressure} J. Guo, H. Wang, F.V. Rohr, Z. Wang, S. Cai, Y. Zhou, K. Yang, A. Li, S. Jiang, Q. Wu, R. J. Cava and, L. Sun, PNAS 114, 13144 (2017).
\bibitem{Cava} K. Stolze, F. A. Cevallos, T. Kong and R. J. Cava, J. Mater. Chem. C 6, 10441 (2018).
\bibitem{muonHEA} K. Motla, Arushi, P. K. Meena, D. Singh, P. K. Biswas, A. D. Hillier, R. P. Singh, Phys. Rev. B 104, 094515 (2021).
\bibitem{rare} J. W. Qiao, M. L. Bao, Y. J. Zhao, H. J. Yang, Y. C. Wu, Y. Zhang, J. A. Hawk, and M. C. Gao, J. Appl. Phys. 124, 195101 (2018).
\bibitem{MF} M. Feuerbacher, M. Heidelmann, and C. Thomas, Mat. Res. Lett. 3, 1 (2015).
\bibitem{noncentrosymmetric} P. A. Frigeri, D. F. Agterberg, I. Milat, and M. Sigrist, Eur. Phys. J. B 54, 435 (2006).
\bibitem{Re6Hf} D. Singh, J.A.T. Barker, A. Thamizhavel, D. McK. Paul, A. D. Hillier, and R. P. Singh, Phys. Rev. B 96, 180501(R) (2017).
\bibitem{Re6Ti} D. Singh, K. P. Sajilesh, J. A. T. Barker, D. McK. Paul, A. D. Hillier, and R. P. Singh, Phys. Rev. B 97, 100505(R) (2018).
\bibitem{MoRe} T. Shang, D. J. Gawryluk, J. A. T. Verezhak, E. Pomjakushina, M. Shi, M. Medarde, J. Mesot, and T. Shiroka, Phys. Rev. Materials 3, 024801 (2019).
\bibitem{Re6Zr} R. P. Singh, A.D. Hillier, B. Mazidian, J. Quintanilla, J. F. Annett, D. McK. Paul, G. Balakrishnan, and M. R. Lees, Phys. Rev. Lett. 112, 107002 (2014).
\bibitem{Re5.5Ta} Arushi, D. Singh, P. K. Biswas, A. D. Hillier, and R. P. Singh, Phys. Rev. B 101, 144508 (2020).

\bibitem{Nb0.5Os0.5} D. Singh, J. A. T. Barker, A. Thamizhavel, A. D. Hillier, D McK Paul, R. P. Singh, Journal of Physics: Condensed Matter 30 075601 (2018).

\bibitem{Re} T. Shang, M. Smidman, S. K. Ghosh, C. Baines, L. J. Chang, D. J. Gawryluk, J. A. T. Barker, R. P. Singh, D. Mck. Paul, G. Balakrishnan, E. Pomjakushina, M. Shi, M. Medarde, A. D. Hillier, H. Q. Yuan, J. Quintanilla, J. Mesot, and T. Shiroka, Phys. Rev. Lett. 121, 257002 (2018).

\bibitem{AuBe} D. Singh, A. D. Hillier, R. P. Singh, Phys. Rev. B 99, 134509 (2019).

\bibitem{NbOs2} D. Singh, Sajilesh. K. P., S. Marik, A. D. Hillier, R. P. Singh, Phys. Rev. B  99, 014516 (2019).

\bibitem{TaOs} D. Singh, Sajilesh K. P., S. Marik, P. K.Biswas, A. D. Hillier, and R. P. Singh, Journal of Physics: Condensed Matter 32 015602 (2020).

\bibitem{Hume} W. Hume Rothery, J. Institute of Metals, 35, 295 (1926).

\bibitem{swavehea} G. Kim, M. -H. Lee, J. H. Yun, P. Rawat, S. -G. Jung, W. Choi, T. -S. You, S. J. Kim, J. -S. Rhyee, Acta Materialia 186, 250 (2020).
\bibitem{carrier} Y -F Cao, S -K Chen, T -J Chen, P -C Chu, J -W Yeh, and S -J Lin, J. Alloy and compd. 509, 1607 (2011). 
\bibitem{Complexity} F. V. Rohr, M. J. Winiarski, J. Tao, T. Klimczuk, and R. J. Cava, Proc. Natl. Acad. Sci. USA 113, E7144 (2016).
\bibitem{whh} N. R. Werthamer, E. Helfand, and P. C. Hohemberg, Phys. Rev. 147, 295 (1966).
\bibitem{maki} A. B. Karki, Y. M. Xiong, N. Haldolaarachchige, S. Stadler, I. Vekhter, P. W. Adams, D. P. Young, W. A. Phelan, and J. Y. Chan, Phys. Rev. B 83, 144525 (2011).
\bibitem{pauli} A. M. Clogston, Phys. Rev. Lett. 9, 266 (1962).
\bibitem{McMillan} W. L. McMillan, Phys. Rev. 167, 331 (1968).
\bibitem{Tinkham}  M. Tinkham, Introduction to Superconductivity (McGraw-Hill, New York, 1996).
\bibitem{U1} Y. J. Uemura, V. J. Emery, A. R. Moodenbaugh, M. Suenaga, D. C. Johnston, A. J. Jacobson, J. T. Lewandowski, J. H. Brewer, R. F. Kiefl, S. R. Kreitzman, G. M. Luke, T. Riseman, C. E. Stronach, W. J. Kossler, J. R. Kempton, X. H. Yu, D. Opie and H. E. Schone, Phys. Rev. B 38, 909 (1988).
\bibitem{U2} Y. J. Uemura, G. M. Luke, B. J. Sternlirb, J. H. Brewer, J. F. Carolan, W. N. Hardy, R. Kadono, J. R. Kempton, R. F. Kiefl, S. R. Kreitzman, P. Mulhern, T. M. Riseman, D. L. Williams, B. X. Yang, S. Uchida, H. Takagi, J. Gopalakrishnan, A. W. Sleight, M. A. Subramanian, C. L. Chien, m. Z. Cieplak, G. Xiao, V. Y. Lee, B. W. Statt, C. E. Stronach, W. J. Kossler and X. H. Yu, Phys. Rev. Lett. 62, 2317 (1989).
\bibitem{U3} Y. J. Uemura, L. P. Le, G. M. Luke, B. J. Sternlieb, W. D. Wu, J. H. Brewer, T. M. Riseman, C. L. Seaman, M. B. Maple, M. Ishikawa, D. G. Hinks, J. D. Jorgensen, G. Saito and H. Yamochi, Phys. Rev. Lett. 66, 2665 (1991).
\bibitem{JAT} J. A. T. Barker, B. D. Breen, R. Hanson, A. D. Hillier, M. R. Lees, G. Balakrishnan, D. McK. Paul and R. P. Singh, Phys. Rev. B 98, 104506 (2018).
\bibitem{Sundar} S. Sundar, S. Salem-Sugui Jr., M. K. Chattopadhyay, S. B. Roy, L. S. Sharath Chandra, L. F. Cohen and L. Ghivelder, Supercond. Sci. Technol. 32 055003 (2018).
\bibitem{Re24Ti5Uemura} C. S. Lue, H. F. Liu, C. N. Kuo, P. S. Shih, J. Y. lin, Y. K. Kuo, M. W. Chu, T. L. Hung, and Y. Y. Chen, Supercond. Sci. Technol. 26, 055011 (2013).
\bibitem{Mayoh} D. A. Mayoh, J. A. T. Barker, R. P. Singh, G. Balakrishnan, D. McK. Paul and M. R. Lees, Phys. Rev. B 96, 064521 (2017).\\

\end{thebibliography}
\end{document}